\documentclass[sigconf]{acmart}
\renewcommand\footnotetextcopyrightpermission[1]{}
\AtBeginDocument{%
  \providecommand\BibTeX{{%
    \normalfont B\kern-0.5em{\scshape i\kern-0.25em b}\kern-0.8em\TeX}}}

\usepackage[ruled,vlined]{algorithm2e}
\usepackage{caption}
\usepackage{subcaption}
\usepackage{mathtools}

\begin{document}

\title{AlphaBlock: An Evaluation Framework for Blockchain Consensus Protocols}

\author{Haitao Xiang}
\email{haitao.xiang@maths.ox.ac.uk}
\orcid{XXXXXXXX}
\affiliation{%
  \institution{Mathematical Institute, University of Oxford}
  \streetaddress{XXX}
  \city{Oxford}
  \state{UK}
  \postcode{XXXXX}
}

\author{Zhijie Ren}
\affiliation{%
  \institution{VeChain}
  \city{Shanghai}
  \country{China}}
\email{zhijie.ren@vechain.com}

\author{Ziheng Zhou}
\affiliation{%
  \institution{VeChain}
  \city{Shanghai}
  \country{China}}
\email{peter.zhou@vechain.com}

\author{Ning Wang}
\email{ning.wang@maths.ox.ac.uk}
\orcid{https://orcid.org/0000-0002-6746-5481}
\affiliation{%
  \institution{Mathematical Institute, University of Oxford}
  \streetaddress{XXX}
  \city{Oxford}
  \state{UK}
  \postcode{OX2 6GG}
}

\author{Hanqing Jin}
\email{jinh@maths.ox.ac.uk}
\orcid{https://orcid.org/0000-0002-6746-5481}
\affiliation{%
  \institution{Mathematical Institute, University of Oxford}
  \streetaddress{XXX}
  \city{Oxford}
  \state{UK}
  \postcode{OX2 6GG}
  }


\begin{abstract}
Consensus protocols play a pivotal role to 
balance security and efficiency in blockchain systems. In this paper, we propose an evaluation framework for blockchain consensus protocols termed as AlphaBlock. In this framework, we compare the overall performance of Byzantine Fault Tolerant (BFT) consensus and Nakamoto Consensus (NC). BFT consensus is reached by multiple rounds of quorum votes from the supermajority, while  NC is reached by  accumulating credibility with the implicit voting from appending blocks. AlphaBlock incorporates the key concepts of Hotstuff BFT (HBFT) and Proof-of-authority (PoA) as the case study of BFT and NC. Using this framework, we compare the throughput and latency of HBFT and PoA with practical network and blockchain configurations. Our results show that the performance of HBFT dominates PoA in most scenarios due to the absence of forks in HBFT. 
Moreover, we find out a set of optimal configurations in AlphaBlock, 
which sheds a light for improving the performance of blockchain consensus algorithms.

\end{abstract}




\maketitle

\section{Introduction}

Blockchain technology is one of the most significant innovations in recent years that may disrupt many industries. As a decentralized and unchangeable ledger that is jointly maintained by untrusted parties without a globally trusted intermediary, it promises integrity, transparency and resilience that are critical to many economic activities \cite{bano2019sok}. Although obtaining its cognition from Bitcoin \cite{nakamoto2019bitcoin}, it has gained attention in more than the financial industry. In government \cite{wons2017digital}, patent \cite{dinizo2018alice}, voting \cite{abraham2016bvp} and real estate \cite{spielman2016blockchain}, blockchain technology serves as an attractive alternative to current centralized intermediaries with huge trust costs.

The main obstacle lying before the wide adoption of blockchain technology is its unsatisfactory performance and poor scalability as compared to its traditional counterparts. This difficult tradeoff between system performance and intermediary trustworthiness stops most businesses interested in blockchain technology. Therefore it is  crucial to find out the features and properties that bottleneck the system performance and scalability. Among the features, the consensus protocol is a critical one. A consensus mechanism guarantees that all honest nodes will eventually agree on a consistency ledger in asynchronous and untrusted networks.

Currently, two types of mainstream consensus protocols prevail. One is Nakamoto Consensus (NC) originated from Bitcoin \cite{nakamoto2019bitcoin}, in which the consensus of a proposed block is not definitive, but probabilistic. More precisely, the probability that there exists another honest node agreeing on a conflicting block drops proportionally to the number blocks appending to it. Then, a block is seen as ``confirmed'' when this probability is lower than a predetermined security parameter. For instance, Bitcoin confirms a block when there are five blocks appending to it. 

The other is Byzantine Fault Tolerant (BFT), in which the consensus is reached through multiple rounds of quorum vote, where a block is guaranteed to reach absolute consistency in the asynchronous network when it receives votes from the supermajority (more than 2/3 of the nodes) two or three times \cite{castro1999practical,yin2019hotstuff}. It is worth noting that NC can be used in either permissioned or permissionless networks. In permissioned networks, the number of honest nodes should be more than 50\% of the population. In permissionless networks, a proof of the possession of a certain resource should be included in each block and the honest nodes should be in possession of more than 50\% of that resource. BFT algorithms, on the other hand, are mostly used in permissioned network where the number of adversaries is less than 1/3 of the population.


Both consensuses have their merits and disadvantages.
BFT algorithms like \cite{castro1999practical} are shown to achieve consensus with higher throughput and lower latency in small networks, yet quorum votes are involved in dampening system performance. NC algorithms \cite{nakamoto2019bitcoin,wood2014ethereum}, on the other hand, does not require quorum votes which is complex in communication. Hence, it is usually believed to achieve higher throughput and lower latency in large practical networks. Recently, new BFT algorithms, like those in \cite{gilad2017algorand,yin2019hotstuff,kwon2014tendermint}, are proposed for blockchains with a relatively large network and good performance under laboratory settings. The design logic of BFT and NC differs in various features including the quorum votes, and these features all affect system performance.
 

Given the different design logic of the two consensuses, researchers and practitioners will question which consensus is better given a certain network condition and blockchain application. Practically, by ``different network conditio'' we mean  different network delay, bandwidth and byzantine ratio. By ``better'' we mean greater throughput and less latency. Therefore, it is desirable to have a framework to compare the performance of two protocol under a practical and general setting.

In literature, there are various evaluation frameworks that compare the performance of blockchain systems. Gervais et al. \cite{gervais2016security} propose a framework to compare different Proof-of-Work (PoW) blockchains. In this framework, the performance of a blockchain is measured by its throughput when a certain security level is obtained under the optimal adversary attack. Dinh et al. \cite{dinh2017blockbench} propose BLOCKBENCH to evaluate the throughput, latency and security of private blockchains. They compare the performance of three major blockchains: Parity, Ethereum and Hyperledger Fabric. They found that these state of the art blockchain systems fall behind a traditional database system for business-level workloads. Croman et al. \cite{croman2016scaling} propose a framework to measure the economic cost of different PoW systems. However, none of the above works ever make the quantitative comparison across the NC and BFT consensuses, needless to say formulating the two consensuses within one coherent framework.

In this paper, we  present AlphaBlock, an evaluation framework to compare the performance of NC and BFT in practice. Critical features of the two consensuses, especially those affecting the performance of the blockchain in practical network scenarios, are modelled, including quorum communication, fork rate, empty block, the interval between blocks, block broadcast etc. Moreover, these features are all translated into two key performance indicators: throughput and latency. We show how network delay, bandwidth, and the ratio of the adversaries could affect the performance of both consensuses. Eventually, we give recommendations and general principles on the selection and the configuration of consensus in various networks. 

\section{Background}
NC lays the critical foundation for not only Bitcoin \cite{nakamoto2019bitcoin}, but also other mainstream cryptocurrency designs like Litecoin \cite{haferkorn2014seasonality} and Ethereum \cite{wood2014ethereum}, which promises a bright future of decentralized payment and financial system \cite{huberman2019economic}. However, with its wide adoption, the concerns over its bottlenecks are getting louder.
The concern of undesirable performance is partially resolved with leader selection based algorithms like \cite{bitcoinng,ouroboros} or graph-based protocol like PHANTOM \cite{sompolinsky2018phantom}. 

The problem of BFT is proposed in the 1980s by Lamport et al. in \cite{lamport} and has been extensively studied before blockchain was born \cite{bracha,benor,castro1999practical}. Essentially, BFT consensus can be categorized into state-machine replication protocols \cite{schneider1990implementing} that finds majority consensus in the presence of malicious agents, which naturally meets the demand of blockchain. Therefore, a lot of BFT based systems were proposed, like Tendermint \cite{kwon2014tendermint}, Algorand \cite{gilad2017algorand}, Casper FFG \cite{buterin2017casper} and BEAT \cite{duan2018beat}. Classical BFT algorithms like PBFT \cite{castro1999practical} are criticized for its poor scalability \cite{vukolic2015quest}, namely, an $\mathcal{O}(N^2)$ message complexity, which is not suitable for blockchains in large networks. Several blockchain application-oriented BFT algorithms like Algorand, Tendermint and Hotstuff BFT (HBFT)\cite{yin2019hotstuff} have reduced the average message complexity per transaction in a normal consensus process to $\mathcal{O}(N)$, which is identical to NC algorithms. In particular, Hotstuff BFT also achieves optimal responsiveness and $\mathcal{O}(N)$ message complexity for view change, which makes it similar in form to NC algorithms.


In the next subsection, we will briefly introduce Proof-of-Authority (PoA) algorithm and HBFT algorithm, which is an NC algorithm and a BFT algorithm, respectively.
In each subsection, we will summarize performance-related factors from 3 levels: network, blockchain and protocol, which will motivate the proposition of our evaluation framework.

\subsection{PoA}

PoA is a type of permissioned NC algorithms that are used by blockchains like Parity\footnote{Proof-of-Authority Chains - Wiki Parity Tech Documentation: https://wiki.parity.io/Proof-of-Authority-Chains} and VeChain\footnote{VeChain: https://www.vechain.com}. In this paper, we propose a simple PoA algorithm which is merely a permissioned version of PoW. Firstly, we assume that honest nodes have a consistent view of Global Standard Time (GST) and the time is divided into rounds with a predetermined duration. Then, associated with each block, rather than a hash preimage proving the computation work conducted by the block proposer, a proof proving that the block is proposed by an authorized node which is in charge of proposing block in that round is provided.

The block will be broadcast to the network through gossip protocol. When each node receives a block, it should first validate the block before broadcast it and append it to its blockchain. Invalid block, i.e., blocks including invalid transactions or proposed by the incorrect or unauthorized nodes, will not be broadcast. A malicious node may create a fork, i.e., more than one valid block in a round. Then, honest nodes should apply the ``longest chain rule'' to determine the valid chain and the valid block. It is guaranteed that honest nodes could eventually reach consensus if blocks could be synchronized in a round and honest nodes are more than 50\% of the network.

\subsection{HBFT}

 In HBFT, in each round, a leader is selected according to a predetermined schedule. When a block is proposed by the leader, the leader will broadcast the block to the whole network and all nodes should send their votes with their signature to the leader to participate in the consensus process. When $2f$ agree with votes from the network are received, a quorum consensus (QC) is reached. Here, $f$ is number of the adversaries and $3f+1\leq N$ holds.
 In HBFT, a three-phase commitment approach will be used to reach consensus for this block. More specifically, a block is only confirmed when it reaches QC for 3 times. In this paper, we consider the chained HBFT proposed in \cite{yin2019hotstuff}, where the three-phase commitment scheme is pipelined in a chain. More precisely, in each round, a block is proposed with $2f+1$ votes from different nodes (including the block proposer). Then, a block is confirmed when it is appended by 2 blocks. 
 Moreover, HBFT could function in partial synchronous network by increasing the timeout every time a QC is failed to be reached before the timeout.
In this paper, in order to make a fair comparison, we consider a synchronous network where the network delay is known and thus the round duration could be set such that $2f$ votes could be collected with high probability.

\section{Framework setup}

As in the background, our framework should consist of a module to describe network properties, a module to describe blockchain properties, and a protocol module that accommodates both BFT and NC consensuses. 
We first introduce factors in the three modules, and define throughput and latency by these factors. Then we introduce  the evaluation framework (AlphaBlock), which is essentially a system with a block proposing leader and a block validating \& confirming committee. We simulate the AlphaBlock multiple times to obtain its average throughput and latency. We also fit BFT and NC into the framework in this section. 

\subsection{Key Concepts}
\label{key concepts}
In this section, we introduce key concepts relating to our AlphaBlock.

\subsubsection{network properties}
\label{netpro}
\hfill\\

\textbf{Number of nodes $N$} 
is the number of agents in the blockchain system. 

\textbf{Committee size $C$} 
is the number of agents of the block committee. The committee is used to validate and confirm a block. In BFT, the committee size $C=N$, namely the whole network needs to reach consensus in validating a block. In NC, a block is validated individually and $C=0$, namely no consensus is needed in validating a block. In AlphaBlock, $C$ can take any integer value between $0$ and $N$, which enable framework to accommodate more consensuses other than BFT and NC. 

\textbf{Byzantine ratio $\alpha$} 
is the ratio of malicious nodes in the blockchain system. The malicious nodes will play to dampen the system performance and lower the system safety. To measure the performance of a system, we consider the case with the worst behaviour of malicious nodes. Therefore, when a block proposing node is malicious, 
we assume it will try to create a fork when we compute fork rate, and it will try to create an empty block when we compute probability of empty block, or conversely block rate.

\textbf{Byzantine number $f$} 
is the number of the malicious nodes, and given $\alpha$ and $N$, we have $f=\lceil (N-1)\alpha\rceil$.

\textbf{Endorsement size $d$} 
is the minimum number of committee members needed to validate a block. In BFT, $d=2f$, namely the leader has to collect at least $d$ votes or signatures to validate a block in a round. In NC, no committee is needed to validate a block, so $d=0$. Another choice of $C$ is also permissible in AlphaBlock.

\textbf{Effective Bandwidth $B_{\textrm{width}}$} 
is the overall network bandwidth determined by 
\begin{equation}
    B_{\textrm{width}}=\frac{\textrm{data  size}}{\textrm{time to broadcast data to the whole network}}.
\end{equation}

\textbf{Communication network $G(N,p,D,B_{\textrm{width}})$} 
is a randomly connected graph with parameters
\begin{itemize}
    \item $N$ nodes;
    \item the linkage probability $p$;
    \item the delay factor $D$;
    \item the effective bandwidth of the network $B_{\textrm{width}}$;
\end{itemize} 
In AlphaBlock, we assume the communication delay $\tau$ between linked nodes is lognormal with mean $0$ and variance $D^2$, and messages propagate along the path with the shortest delay. 

\subsubsection{blockchain properties}
\hfill\\

\textbf{Block size $B_{\textrm{size}}$} 
is the maximum number of transactions a block can hold. Since we are interested in the system theoretical performance, in the AlphaBlock workflow we assume each time a block is proposed, it contains $B_{\textrm{size}}$ transactions.

\textbf{Message size $M_{\textrm{size}}$} 
is the size of a message. We define the message as information other than block, because the propagation of these information is dominated by network latency rather than bandwidth.

\textbf{Security level $\epsilon$} 
is a maximal acceptable level of 
the probability that an inconsistency occurs in a system. Here, an inconsistency means that two honest nodes agree on different blocks. 
It seems that 
AlphaBlock is unfair for BFT in the sense that 
the consistency in BFT is definitive.  However, as it is commonly believed that a small enough $\epsilon$ is secure enough and NC and BFT are indistinguishably used in many blockchain applications, it is reasonable to directly compare the performance of NC and BFT giving a small $\epsilon$.



\textbf{Simulation rounds $SR$} 
is the number of rounds of simulation of AlphaBlock to compute the system performance. 

\subsubsection{protocol properties}
\hfill\\

\textbf{Block interval $B_{\textrm{interval}}$} 
is the time interval between two blocks. It is subject to three factors: block broadcast time (BBT), committee communication time (CCT), and blockhead broadcast latency (BBL). The BBT is the time needed to broadcast the block to the whole network, expressed as:
\begin{equation}
\textrm{BBT} = B_{\textrm{size}}/B_{\textrm{width}}.
\end{equation}
The CCT is consensus-based. In BFT and all AlphaBlock with a non-zero endorsement committee, it takes 1 round of communication for the leader to collect votes and send out collected votes, and the time is delay dominant, so approximately two times the $d^{th}$ delay.
\begin{equation}
\textrm{CCT}_{C>0} = 2\times d^{th} \textrm{ delay from leader to committee}.
\end{equation}
In NC, it does not require committee communication, so \begin{equation}
\textrm{CCT}_{C=0} = 0.
\end{equation}
The BBL is the biggest delay of block broadcast, so it is the maximum of delay from the leader to all $N$ member nodes.
\begin{equation}
\textrm{BBL} = \max(\textrm{delay from leader to } N \textrm{ members}).
\end{equation}
Putting together, we have
\begin{equation}
    B_{\textrm{interval}}=\max(\textrm{BBT},\textrm{CCT},\textrm{BBL}).
\end{equation}

\textbf{Fork rate $F_{\textrm{rate}}$} 
is the probability of fork. In AlphaBlock, a fork emerges when 
the  leader and   no less than $d$ committee members are malicious. Therefore
\begin{equation}
\label{eq:FR}
F_{\textrm{rate}} = Pr(\textrm{fork}) = \alpha\times \textrm{Hygcdf}(d|N-1,N\alpha-1,C).
\end{equation}

where $\textrm{Hygcdf}(X|M,K,Y)$ is the cumulative distribution function of hypergeometric distribution as follows:
\begin{equation}
     \textrm{Hygcdf}(X|M,K,Y) = \sum_{0\leq x < X} \frac{{K \choose x}{M-K \choose Y-x}}{{M \choose Y}}.
\end{equation}
Interestingly, equation \ref{eq:FR} holds in both BFT and NC. In BFT, the fork rate is 0, so equation \ref{eq:FR} applies. In NC, the fork rate is $\alpha$, so equation \ref{eq:FR} applies as well. 

\textbf{Block rate $B_{\textrm{rate}}$} 
is the probability of successfully proposing a block in a round, which requires no compromise on liveness. We compute block rate by considering the worst case attack on liveness, as shown in Table \ref{tab:liveness},  which means when a committee could either create a fork or empty round, it would choose an empty round. Therefore
\begin{equation}
\label{eq:BR}
B_{\textrm{rate}}=(1-\alpha)\times(1-\textrm{Hygcdf}(C-d+1|N-1,N\times\alpha,C)).
\end{equation}
Interestingly, equation \ref{eq:BR} holds in both BFT and NC. In BFT, the block rate is $1-\alpha$, so equation \ref{eq:BR} applies. In NC, the fork rate is $1-\alpha$, equation \ref{eq:BR} applies as well. 
\begin{table}
  \caption{Worst case attack on liveness}
  \label{tab:liveness}
  \begin{tabular}{p{1cm}p{1cm}p{2cm}p{3cm}}
    \toprule
    Case&Leader&Committee member&Probability\\
    \midrule
    Case 1&Malicious&\#$\geq$0&$\alpha$\\
    Case 2&Good&\# of good$<$d&$(1-\alpha)\times \textrm{Hygcdf}(C-d+1,N-1,N\times\alpha,C)$\\
  \bottomrule
\end{tabular}
\end{table}

\textbf{Committee rate $C_{\textrm{rate}}$} 
is the proportion of bandwidth used by committee communication.
\begin{equation}
C_{\textrm{rate}} = \frac{\# \textrm{of CC edges}\times M_{\textrm{size}}}{\# \textrm{of CC edges} \times M_{\textrm{size}}+\# \textrm{of broadcast edges}\times B_{\textrm{size}}},
\end{equation}
where $CC$ means committee communication. 
In BFT, $C_{\textrm{rate}}$ is non-zero but small. In NC, $C_{\textrm{rate}}$ is 0.

\textbf{Bandwidth efficiency $B_{\textrm{eff}}$} 
is how efficiently the bandwidth is used for the transmission of the eventual confirmed transactions. Two relevant factors: fork will waste bandwidth, and committee communication will also occupy bandwidth. For simplicity, we first consider the scenario that the malicious leader and committee will create exactly one fork when they have the chance to do so, and consider other scenarios in Subsection~\ref{ss:fr}. Subtracting the two factors will give net block propagating bandwidth. Therefore, bandwidth efficiency goes as follows:
\begin{equation}
B_{\textrm{eff}}=(1-F_{\textrm{rate}})\times(1-C_{\textrm{rate}}).
\end{equation}

\textbf{Confirmation number $K$} is the number satisfying 
\begin{equation}
(F_{\textrm{rate}})^K\leq\epsilon,
\end{equation}
where $\epsilon$ is the security level. 

Note that the confirmation of HBFT and PoA are very different as the consensus types and the synchronous assumptions are different. However, the performance of both algorithms are compared directly regardless of their consensus. Hence, in a practicality perspective, we simply use ``double standards'' on the confirmation, according to their own confirmation rules in each consensus algorithm.

\begin{equation}
   K = \left\{
\begin{aligned}
\lceil \log{\epsilon}/\log F_{\textrm{rate}} \rceil, & & \textrm{PoA}\\
3, & & \textrm{HBFT}
\end{aligned}
\right..
\end{equation}

\subsubsection{System Performance}
\hfill\\

We are interested in the throughput and latency of the blockchain system. 

\textbf{Throughput $T$} 
is evaluated as transactions per second, and it is originally concerned with number of transactions within a block interval. However, due to forks, empty blocks, and waste of bandwidth due to committee communication, the throughput at security level $\epsilon$ is adjusted as follows:
\begin{equation}
    T=\frac{B_{\textrm{size}}}{B_{\textrm{interval}}}\times B_{\textrm{rate}}\times B_{\textrm{eff}}.
\end{equation}

\textbf{Latency $L$} 
is evaluated as average time delay to confirm a transaction. it is originally concerned with number of appending blocks $K$ needed to confirm a block, which is originally equivalent to number of block intervals. However, due to empty rounds, $K$ needs to be adjusted by block rate, so the latency at security level $\epsilon$ is as follows:
\begin{equation}
    L=\frac{K}{B_{\textrm{rate}}}\times B_{\textrm{interval}}.
\end{equation}

\begin{figure*}
\centering
 \begin{subfigure}[b]{0.3\textwidth}
         \centering
         \includegraphics[width=\textwidth]{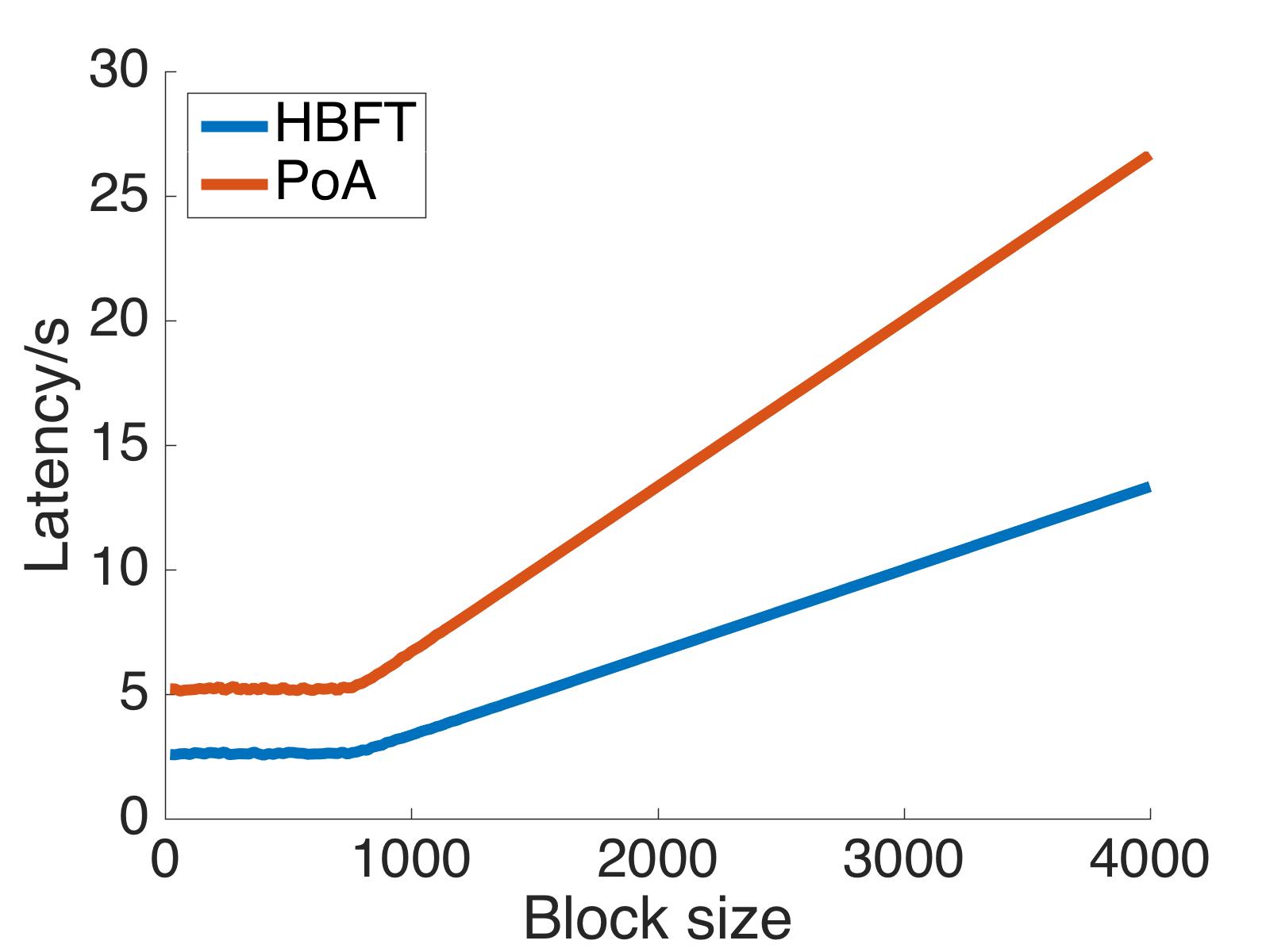}
         \caption{Block size}
             \label{fig1a}
     \end{subfigure}
\begin{subfigure}[b]{0.3\textwidth}
         \centering
         \includegraphics[width=\textwidth]{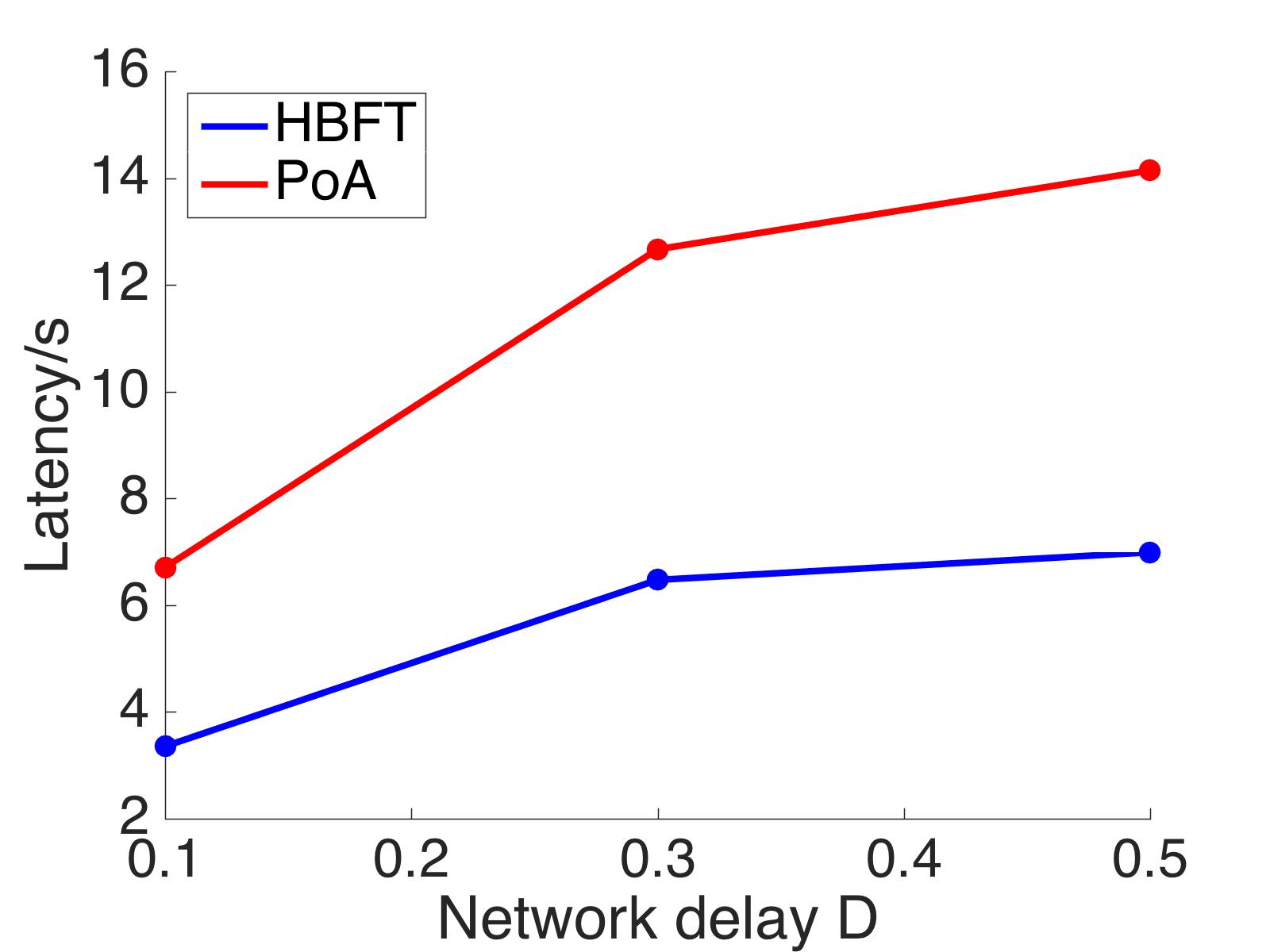}
         \caption{Network delay factor $D$}
         \label{fig1b}
     \end{subfigure}
\begin{subfigure}[b]{0.3\textwidth}
         \centering
         \includegraphics[width=\textwidth]{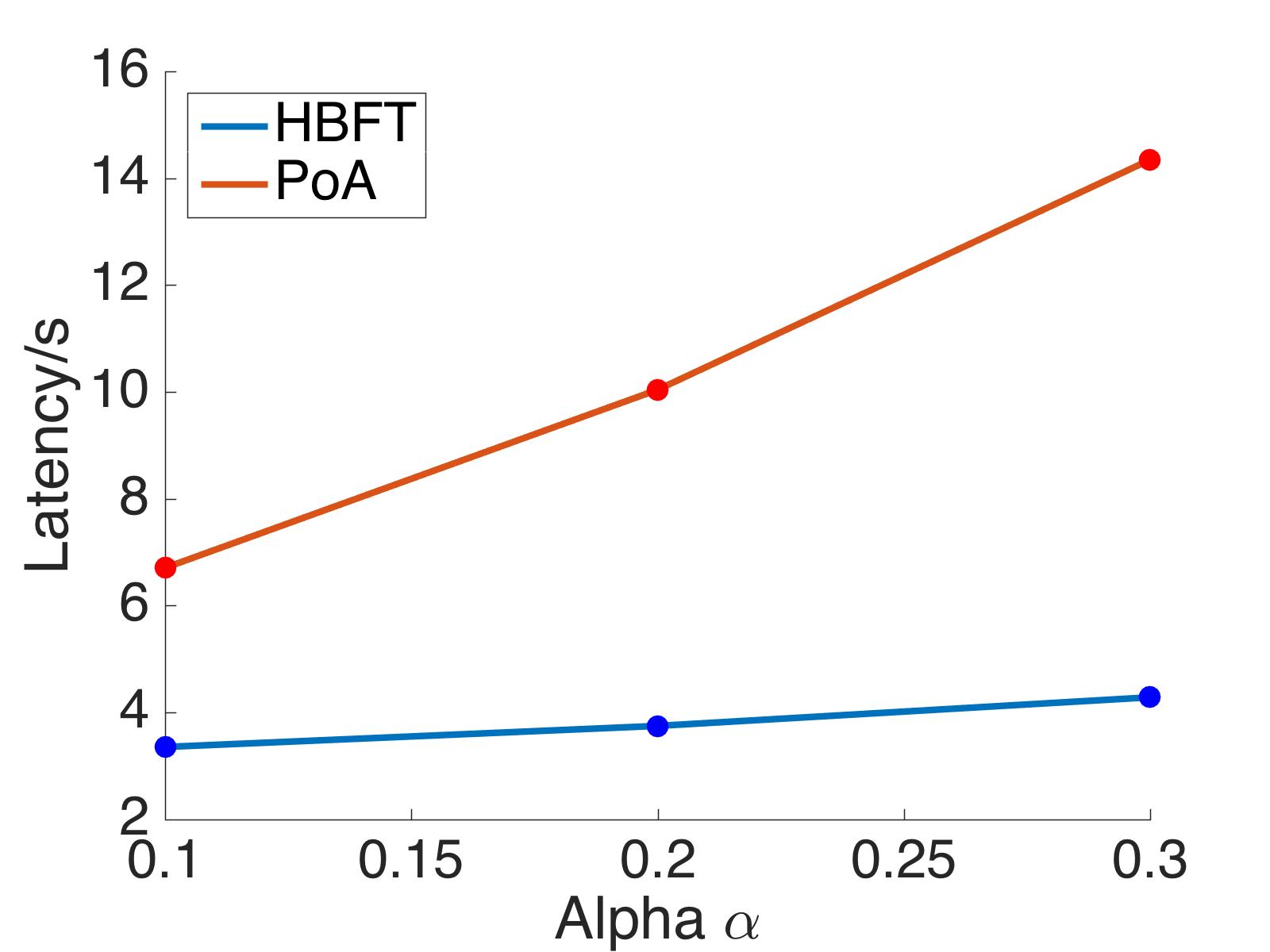}
         \caption{Alpha $\alpha$}
         \label{fig1c}
     \end{subfigure}
\caption{BFT vs. NC latency as a function of block size, network delay, and Byzantine ratio.}
\label{fig1}
\end{figure*}

\subsection{AlphaBlock}

With key concepts defined, we propose the evaluation framework, namely AlphaBlock, which is able to accommodate to both PoA of NC and HBFT of BFT. AlphaBlock loop over the following five parts adapted from \cite{xiao2020survey}: Initialization; block proposal; information propagation; block validation; block finalization.
\begin{itemize}
    \item \textbf{Initialization}. A communication network $G(N,p,D,B_{\textrm{width}})$ is set according to \ref{netpro}, in which there are a proportion of $\alpha$ malicious or byzantine nodes. At the beginning of each round, A leader is randomly selected to propose a block, which matches PoA and HBFT. A committee of size $C$ is randomly selected to validate a block, and at least $d$ endorsements are needed to validate a block. In HBFT, $C=N, d=2f$. In PoA, $C=0,d=0$. 
    \item \textbf{Block Proposal}. In a round, the leader proposes a block, in which $B_{\textrm{size}}$ transactions are added by the leader. The probability that selected leader happens to be malicious is $\alpha$. If the leader is malicious, it may create an empty block, and the probability of a non-empty block is  block rate $B_{\textrm{rate}}$. The malicious leader may also create a fork, and the probability of a fork is fork rate $F_{\textrm{rate}}$. Note $B_{\textrm{rate}}$ and $F_{\textrm{rate}}$ apply to both HBFT and PoA as explained in key concepts.
    \item \textbf{Information propagation}. When a block is proposed by the leader, the leader will send the block to the whole network using gossip, which takes time the bigger of broadcast time BBT and BBL. Meanwhile the leader communicate with the committee to confirm the block, which takes time CCT. Note BBT, BBL and CCT are explained in block interval of key concepts and all accommodated to HBFT and PoA. Therefore, the block propagation time BBT, the block broadcast latency BBL, and the back-and-forth of committee communication time CCT should all be smaller than or equal to block interval to ensure successful propagation. 
    \item \textbf{Block validation}. After Each node receiving the block, AlphaBlock will need to implement $R$ rounds of votes to validate the block. When the block is validated, it is appended to the blockchain, and the round starts over. In each round, at least $d$ votes must be collected to validate the block, otherwise the block will be nullified and the round will fail. In HBFT, a node will agree the proposed block by signing it and sending the signed vote to the leader. The leader will collect at least $2f$ signatures sent by other nodes, wrap them up, and broadcast the wrapped collected signatures to the whole network. Nodes then add the received wrapped signatures into the block before appending the block to its local blockchain; In PoA, it will take $R=0$ round: a node will just validate the block as long as the leader's signature is verified, and there is no need of extra committee communication.
    \item \textbf{Block confirmation}. To confirm a transaction in a block, It requires $K-1$ appending blocks. As explained in the key concepts, $K$ is protocol based. In HBFT, $K=3$, namely it takes 3 rounds of votes from the super-majority to achieve definitive consensus. In PoA, $K=5$, and it only guarantees a probabilistic consensus with a security level of $\epsilon$. 
\end{itemize}

We present the above idea in Algorithm \ref{algo:alg1}. As a simplified version, instead of validating a block, we compute the exact probability of a block being validated and compute variables regardless of the validity of the round. This is reasonable because the variables computed within and without a valid round only differ with respect to the Byzantine ratio, but the communication network is assumed symmetric with respect to being Byzantine. The schematic workflow goes as in Algorithm \ref{algo:alg1}:
\begin{algorithm}
\SetKwInput{KwInput}{Input}               
\SetKwInput{KwOutput}{Output} 
\SetAlgoLined
\KwInput{A random connected graph $G(N,p,D,B_{\textrm{width}})$; A blockchain model $Ch(u,r=0)$, where $r$ is the \# of valid round, and $u$ is the index of the leader in round $r$ that succeeds to generate a block; parameter $C$, $d$, $\alpha$, $\epsilon$ , $B_{\textrm{size}}$, $SR$.}
\KwOutput{$(T(C,d,B_{\textrm{size}}),L(C,d,B_{\textrm{size}}))$, namely Throughput-Latency pair as a function of $C,d$, and $B_{\textrm{size}}$.}
\For{$i = 1$ to $SR$}
{ $Com=randint(N,C+1)$
\tcp*[l]{(Initialization)}
$r=com(0), \textrm{broadcast block }B_{r}\textrm{ to }com(j),j=1,...C$
\tcp*[l]{(proposal)}
$com(j)\textrm{ vote or not to the leader }r$
\tcp*[l]{(propagation)}
$\textrm{Broadcast to all } N \textrm{ nodes to append } B_{r} \textrm{ to } Ch(u,r), r=r + 1$;
\tcp*[1]{(validation, confirmation)}
$\textrm{Compute the variables in Section} \ref{key concepts} \textrm{ except } T \textrm{ and }\ L$
}
$\textrm{compute\ expected\ block\ interval } B_{\textrm{interval}} \textrm{ and committee rate } C_{\textrm{rate}}$;
$\textrm{compute } T \textrm{ and } L \textrm{ with expected } B_{\textrm{interval}} \textrm{ and } C_{\textrm{rate}}$;

\Return{$T,L$}
 \caption{Committee based consensus algorithm}
 \label{algo:alg1}
\end{algorithm}

\subsection{Frontier Identification Algorithm}
AlphaBlock accommodates more protocols than HBFT and PoA. 
It is interesting to find some optimal protocols in this framework by changing controllable parameters. 
In terms of throughput and latency, we can figure out 
these optimal protocols by the throughput-latency frontier
in the sense that 
the throughput cannot be improved without compromising latency and vice versa. 
The frontier is a strong guidance of future blockchain system design. 

More accurately, we are interested in  optimising the throughput and latency of a system by choosing proper parameters $(C,d,B_{\textrm{size}})$ from a certain feasible set ${\mathbb D}$. If we denote by $T(C,d,B_{\rm size})$ and $L(C,d,B_{\rm size})$  the throughput and latency of the system with parameters $(C,d,B_{\rm size})$, then the problem can be formulated as 
\begin{equation}
    \min{-T(C,d,B_{\textrm{size}})}, \min{L(C,d,B_{\textrm{size}})}, \ (C,d,B_{\textrm{size}})\in {\mathbb D}.
\end{equation}
The solution of this two-objective optimisation can be defined by 
\emph{Pareto Dominance}. 
Given $y^{1}=(-t^{1},l^{1})$ and $y^{2}=(-t^{2},l^{2})$ then $y^1$ is said to \emph{Pareto Dominate} $y^{2}$(namely $ y^{1} \prec_{Pareto} y^{2}$), iff
\begin{equation}
    \forall i\in{1,2}: y_{i}^{1}\leq y_{i}^{2}\ and\ \exists j\in{1,2}:y_{j}^{1}< y_{j}^{2}.
\end{equation}
and the set of all optimal solutions to the problem is described as the  \emph{Pareto Frontier}, which is defined as
\begin{equation}
P(Y)=\{y'\in Y:\{y''\in Y:y''\prec_{Pareto} y',y''\neq y'\}=\emptyset\}.
\end{equation}
Namely, Pareto Frontier is the set in which no other points outside the set can Pareto dominate any point in the set. We find the Pareto frontier by Algorithm 2.
\begin{algorithm}
\SetKwInput{KwInput}{Input}               
\SetKwInput{KwOutput}{Output} 
\SetAlgoLined
\KwInput{Y=(-T,L)}
\KwOutput{Pidx ,namely Index for Pareto Frontier of Y.}
$[~,idx] = sort(T,'ascend')$;

\For{$i = 1$ to length(idx)}{$[I] = find(L==min(L (idx(i:end))),1)$;

$Lidx(i) = I$;}
$Lidx = unique(Lidx);$

$[~,idx] = sort(L,'ascend')$;

\For{$i = 1$ to length(idx)}{$[I] = find(T==min(T (idx(i:end))),1)$;

$Tidx(i) = I$;}
$Tidx = unique(Tidx);$

$Pidx = intersect(Lidx,Tidx)$

\Return{Pidx}
 \caption{Frontier identification algorithm}
\end{algorithm}

\section{Results and discussion}

In this section, unless otherwise specified, we use the baseline parameter setting specified in Table 2 . Furthermore, the Byzantine ratio is assumed to be 0.1, which is mild. The security level $\epsilon$ is taken so that the confirmation number $\emph{K}$ of PoA is 5. We first choose the number of agents to be $N=101$ as seen in VeChain and Libra\footnote{https://libra.org/en-US/white-paper/?noredirect=en-US}. The connection probability $\emph{p}$ is taken to be 0.06, which is typical to model internet \cite{boccaletti2006complex}. The network delay parameter is fitted from the data in \cite{gencer2018decentralization}. The message size is 
approximated from \cite{georgiadis2019many}. The bandwidth is a mild assumption adapted from \cite{croman2016scaling}. The block size is assumed from typical data of Bitcoin \cite{georgiadis2019many}.
\begin{table*}
  \caption{parameter setup}
  \label{tab2}
  \begin{tabular}{p{1cm}p{1cm}p{1cm}p{1cm}p{2cm}p{2cm}p{2cm}p{2cm}}
    \toprule
    $\alpha$&$\epsilon$&$N$&$p$&Networkdelay factor $D$&Message size $M_{\textrm{size}}$&Bandwidth $B_{\textrm{width}}$&Block size $B_{\textrm{size}}$\\
    \midrule
    0.1&$10^{-5}$&101&0.06&0.1&0.1 KB&1 MB/s&2000 tx/block\\
  \bottomrule
\end{tabular}
\end{table*}

\subsection{Latency}

In this subsection, we focus on the latency of HBFT and PoA with different values of block size, network delay and Byzantine ratio, respectively.
We take the block size $B_{\textrm{size}}\in \{1,2,...100\}\times40$, the network delay $D\in \{0.1,0.3,0.5\}$, and the byzantine ratio  $\alpha \in \{0.1,0.2,0.3\}$. We show the dependence of the latency on these three variables and the comparison between HBFT and PoA by our numerical experiments shown in Figure \ref{fig1}.

In the left panel of Figure 1, we can see that the latency of both consensus protocols remain flat when the block size stays below 1000, and start to rise after the block size exceeds 1000. This phenomenon can also be seen from the definition of the latency, in which the block size plays it role through the block interval defined as the maximum value among 
$B_{\textrm{size}}/B_{\textrm{width}}$, the committee  communication, and the blockhead broadcast. This maximum will increase together with the increasing of block size only when the block size is big enough, in which case the increasing is a linear style. Furthermore, with the bigger confirmation number $K$,  the slope of this increasing for PoA should be bigger, and this is also shown in Figure 1. By the definition of latency, block size only affects latency through block interval, the behavior of latency regarding block size results from block interval. In fact, the flat region is mainly due to the block interval setting that takes maximum from three variables as in key concepts. When block size is smaller than around 1000, the block interval will remain unchanged because the blockhead broadcast and committee communication are dominant to the determination of block interval. Therefore, we see a flat region. When block size larger than 1000, the block interval is dominated by block broadcast, so latency increases linearly in block size. The slope of PoA is greater than HBFT because of their difference in confirmation number $K$.

The middle panel of Figure 1 shows that the latency is increasing in the networkdelay factor $D$ for both HBFT and PoA. 
This monotonicity can also be explained by the block interval. 
For a given block size, a bigger network delay factor $D$ will result in a bigger 
blockhead broadcast and a bigger committee communication, which pushes the latency higher. While our figure shows that this increasing relation is strict, it is true only for big block size according to our explanation here. 

The right panel of Figure \ref{fig1} shows that latency is increasing in Byzantine ratio for both HBFT and PoA. For PoA, this monotonicity comes from the increasing confirmation number $K$ decreasing block rate, as implied in Equation (9) and (13). While for HBFT, this monotonicity comes only from the  decreasing block rate. Therefore, it is natural that PoA has a greater slope of the increasing than HBFT, which is also confirmed in Figure \ref{fig1}. 

Compare the latency between HBFT and PoA in all three panels in Figure \ref{fig1},  we can find that  HBFT has less latency than PoA, and we interpret this comparison mainly by the higher confirmation number $K$ of PoA. While this conclusion 
{\it contradicts the common belief} that BFT performs poorer than PoA when the block size is small. The key argument in this common belief is that BFT has a heavier committee communication overhead. This argument is true for traditional BFT whose communication overhead is $\mathcal {O}(N^2)$. While for HBFT, the communication overhead is reduced to  $\mathcal {O}(N)$, and hence the argument does not apply. 

\subsection{Throughput}

\begin{figure*}
\centering
 \begin{subfigure}[b]{0.3\textwidth}
         \centering
         \includegraphics[width=\textwidth]{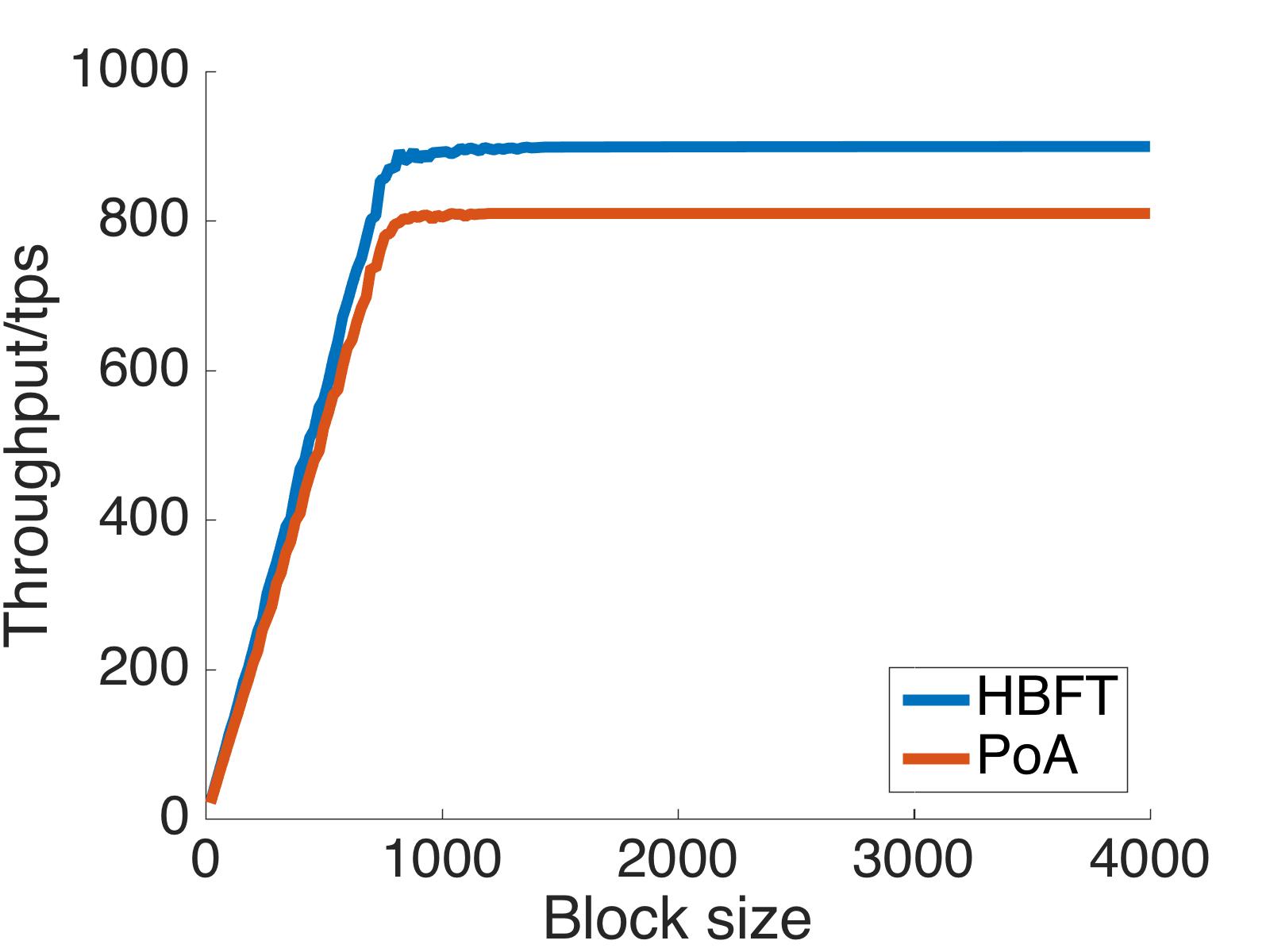}
         \caption{Block size}
             \label{fig2a}
     \end{subfigure}
\begin{subfigure}[b]{0.3\textwidth}
         \centering
         \includegraphics[width=\textwidth]{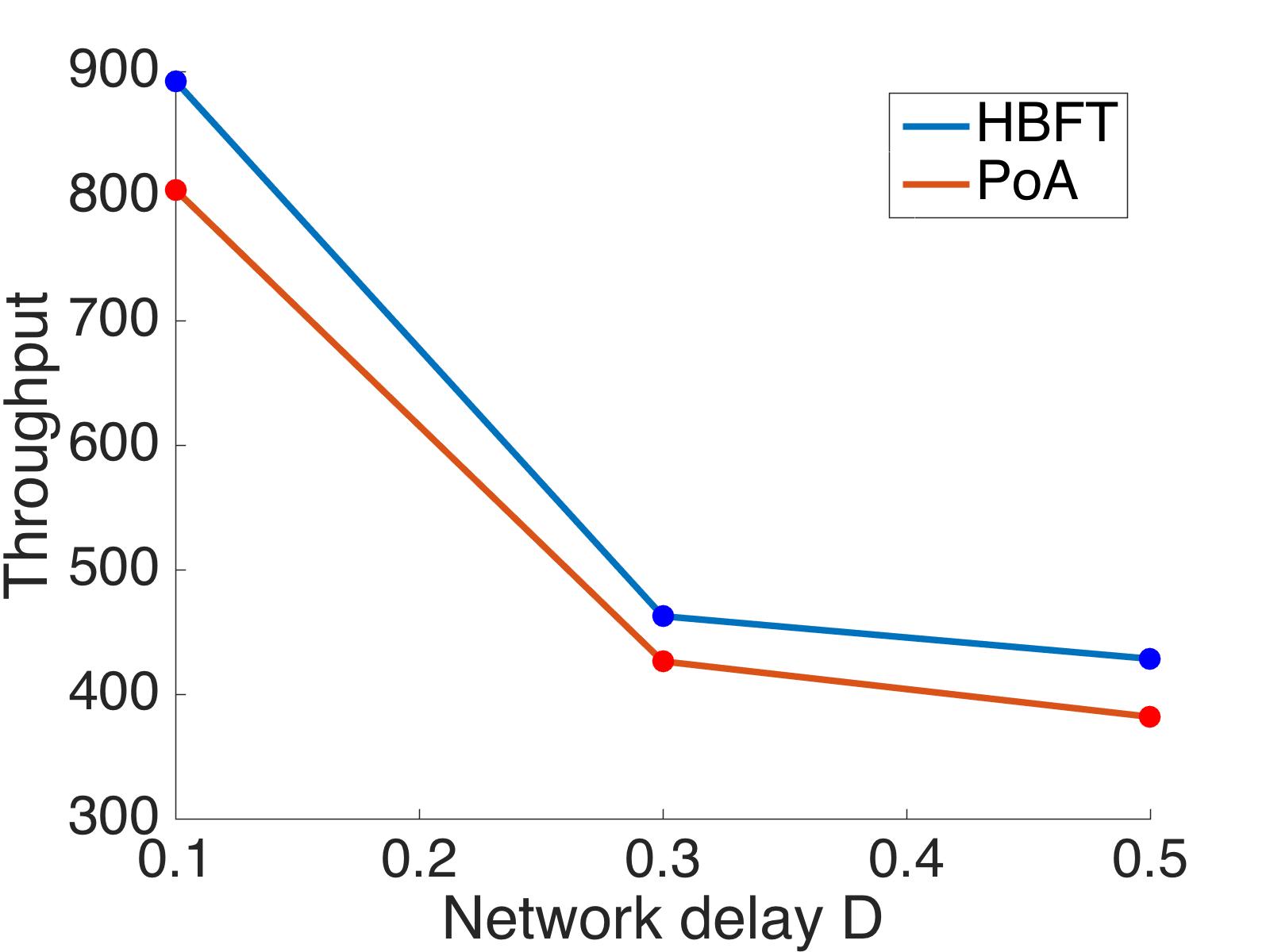}
         \caption{Network delay factor $D$}
         \label{fig2b}
     \end{subfigure}
\begin{subfigure}[b]{0.3\textwidth}
         \centering
         \includegraphics[width=\textwidth]{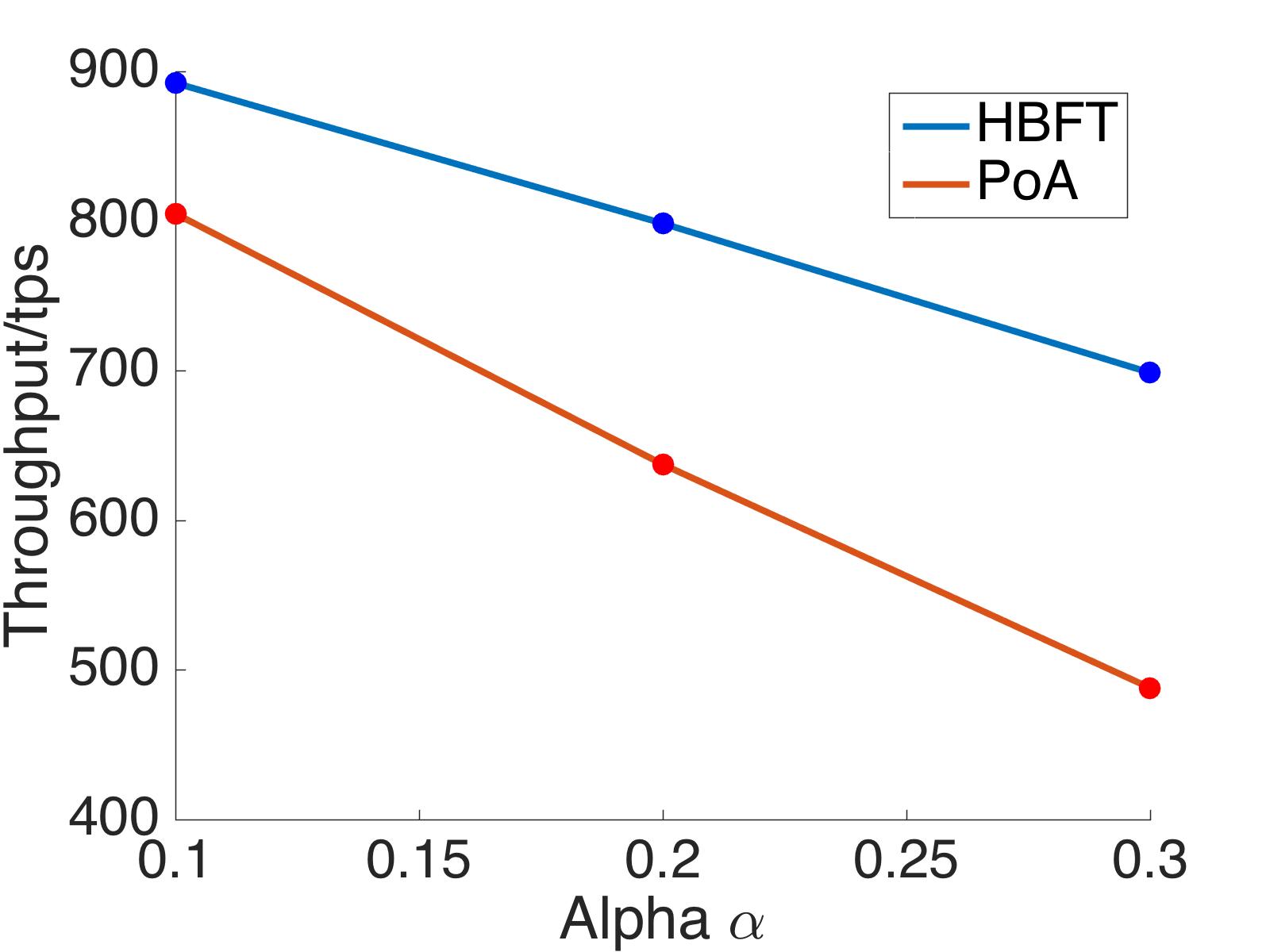}
         \caption{Alpha $\alpha$}
         \label{fig2c}
     \end{subfigure}
\caption{HBFT vs. PoA throughput as a function of block size, network delay, and Byzantine ratio.}
\label{fig2}
\end{figure*}

\begin{figure*}
\centering
\includegraphics[width=18cm]{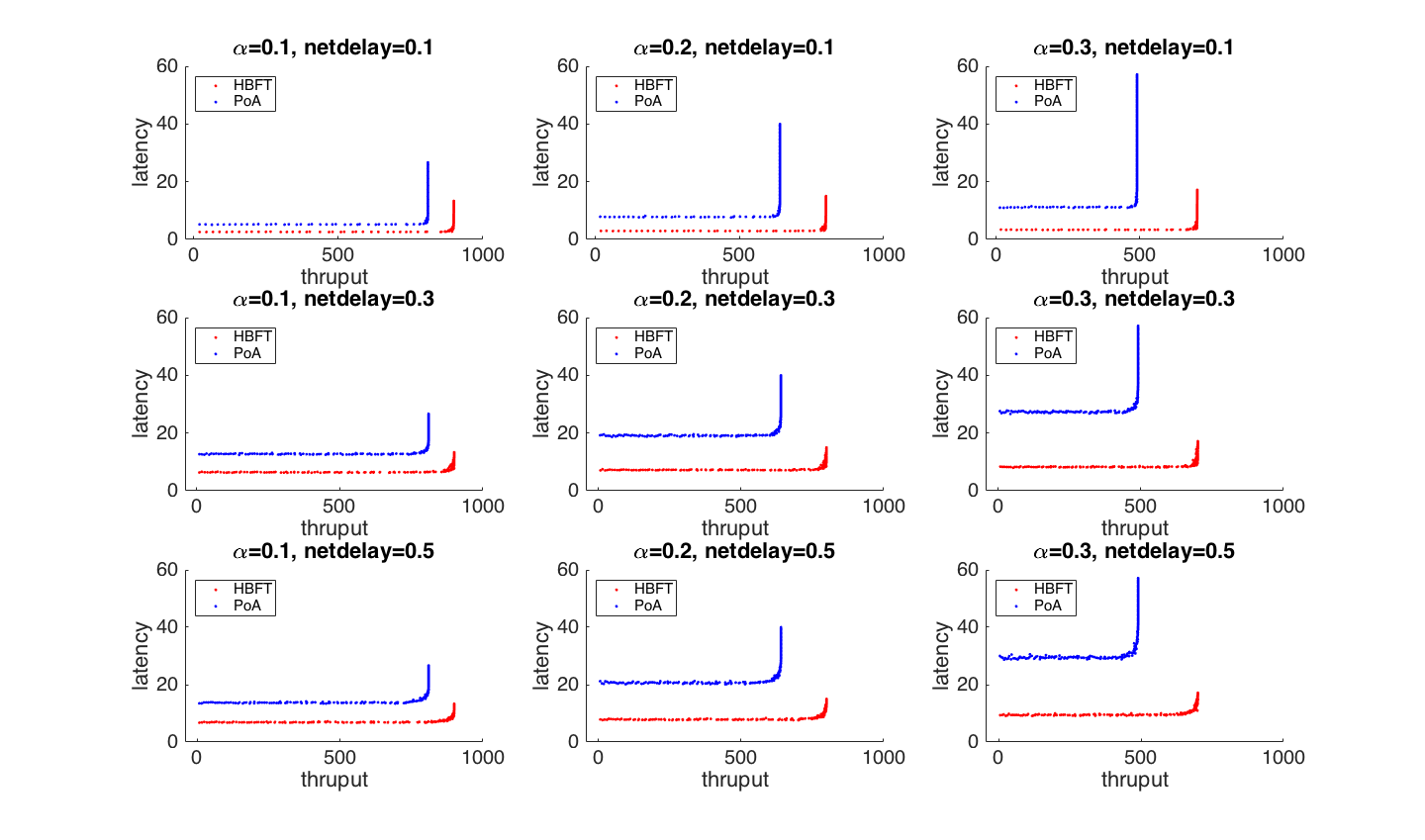}
\caption{HBFT vs. PoA throughput-latency plot across different byzantine ratio $\alpha$ and network delay factor $D$, and the block size range from 40 to 4000.}
\label{fig3}
\end{figure*}

In this subsection, we study the throughput of HBFT and PoA as a function of the  block size, the network delay and the Byzantine ratio.  We take the block size 
$B_{\textrm{size}}\in \{1,2,...100\}\times40$, the  network delay factor $D$ in the set  $D\in \{0.1,0.3,0.5\}$, and the Byzantine ratio  $\alpha \in \{0.1,0.2,0.3\}$. Our experimenal results are shown in Figure \ref{fig2}. To understand implications of these figures, we need the following theorem, whose proof is  deferred to Appendix.
\begin{theorem}
\label{thm}\rm
 As the block size $B_{\textrm{size}}\rightarrow\infty$,  $$\frac{\mbox{Throughput of PoA}}{\mbox{Throughput of HBFT}} \rightarrow (1-\alpha).$$
\end{theorem}
In fact, the limit in Theorem 1 is the ratio of $1-\mbox{Fork rate}$ between the two consensuses, and we know the fork rate of PoA is $\alpha$, while it is $0$ for HBFT.

In 
Figure \ref{fig2a}, we can see that 
the throughput of both consensus protocols increase linearly when the block size is below 1000, and then remain flat after the block size exceed 1000. 
We can examine this phenomenon by the definition of the  throughput as in (13), in which the block rate $B_{\textrm{rate}}$ is a constant, the bandwidth efficiency $B_{\textrm{eff}}$ is very close to the 
constant value $1-F_{\textrm{rate}}$. So the throughput is  approximately a linear function of $B_{\textrm{size}}$. As explained in the Subsection 4.1, the block interval remains a constant when  When the block size is below 1000, and  increases linearly 
otherwise. This shows the dependence of the throughput on the block size. With the flat property and the limit in Theorem 1, we can conclude that the throughput of PoA is lower than that of HBFT when the block size is big. This conclusion can also be interpreted in another way: when the blockchain system is dominated by bandwidth bottleneck, PoA throughput will be smaller than HBFT throughput. 


Figure \ref{fig2b} shows then moving of throughputs when the network delay factor $D$ changes. 
We can see that  both throughputs 
are decreasing in $D$.
The reason is similarly 
due to changes in the block interval, as analyzed in the last subsection. Moreover, we can see from the figure that 
BFT throughput is more sensitive to the network delay. 
This is consistent with the comparison of the block intervals. The committee communication is $0$ in PoA, but it can be the dominating factor among the three factors involved in the block interval, which makes the throughput of HBFT more sensible. 

In Figure \ref{fig2c}, we can see that throughput of both HBFT and PoA are decreasing in the Byzantine ratio $\alpha$. This can be confirmed by the fact that  both the block rate and the fork rate are decreasing in $\alpha$.

In all three panels of Figure \ref{fig2}, 
we find that 
HBFT has  better throughput than PoA. Although this is not an unconditional conclusion, we are confident that it holds for most practical settings of parameters. 

\subsection{Throughput-latency performance plot}

Throughputs and latencies are studied separately in the last two subsection. In this subsection, we check these two measures jointly for PoA and HBFT in different settings. In Figure \ref{fig3}, we present the joint performance on 
throughtputs and latencies for PoA and HBFT with different 
parameter settings.
 In all 9 settings of parameters,  we find  that HBFT dominates PoA in the sense that HBFT has higher throughput and lower latency.  
In this dominance, the higher fork rate of PoA plays a key role, which pushes up the latency and pulls down the throughput. 
Furthermore, in all cases, when the throughput is pushed up, the latency stays approximately  flat before some critical value of throughput, and then spikes up a little bit the value. For a throughput-latency curve,  we call the point on the curve with the critical throughput as the turning point (Tpt) of the curve. A Tpt is informative to measure the performance of a consensus protocol, in which  
 its throughput and latency  indicate  the bottleneck
 values of the protocol. 
 In all $3\times3$ panels in Figure \ref{fig3}, we have the following observations. 
\begin{itemize}
    \item Tpt's throughputs for both HBFT and PoA decrease in $\alpha$. This naturally arise from the block rate in bandwidth efficiency. The difference between the Tpt's throughputs of HBFT and PoA increases in $\alpha$ because of the difference in their fork rates. 
    \item Tpt's throughput for both PoA and HBFT remains almost unchanged in the network delay factor $D$. This means that the network delay has little impact on the bottleneck throughput of either system.
    \item Tpt's latency for PoA increases in $\alpha$, but  remains unchanged for HBFT, 
    so their difference increases in $\alpha$. This phenomenon can be confirmed by the fact that 
     the  fork rate for HBFT does not depend on  $\alpha$, but it equals $\alpha$ for PoA. We conclude that the Byzantine ratio affects the bottleneck latency for PoA only.
    \item Tpt's latency for PoA and HBFT increases in the network delay factor $D$. This is natural given the fact that  a greater delay leads to a greater block interval in general. The difference between the two Tpt's latencies 
    increases in the network delay factor $D$, which is due to the greater  confirmation number $K$ of PoA compared with that of  HBFT.
\end{itemize}

\subsection{Alternative fork strategy}\label{ss:fr}
\begin{figure}
\centering
\includegraphics[width=8cm]{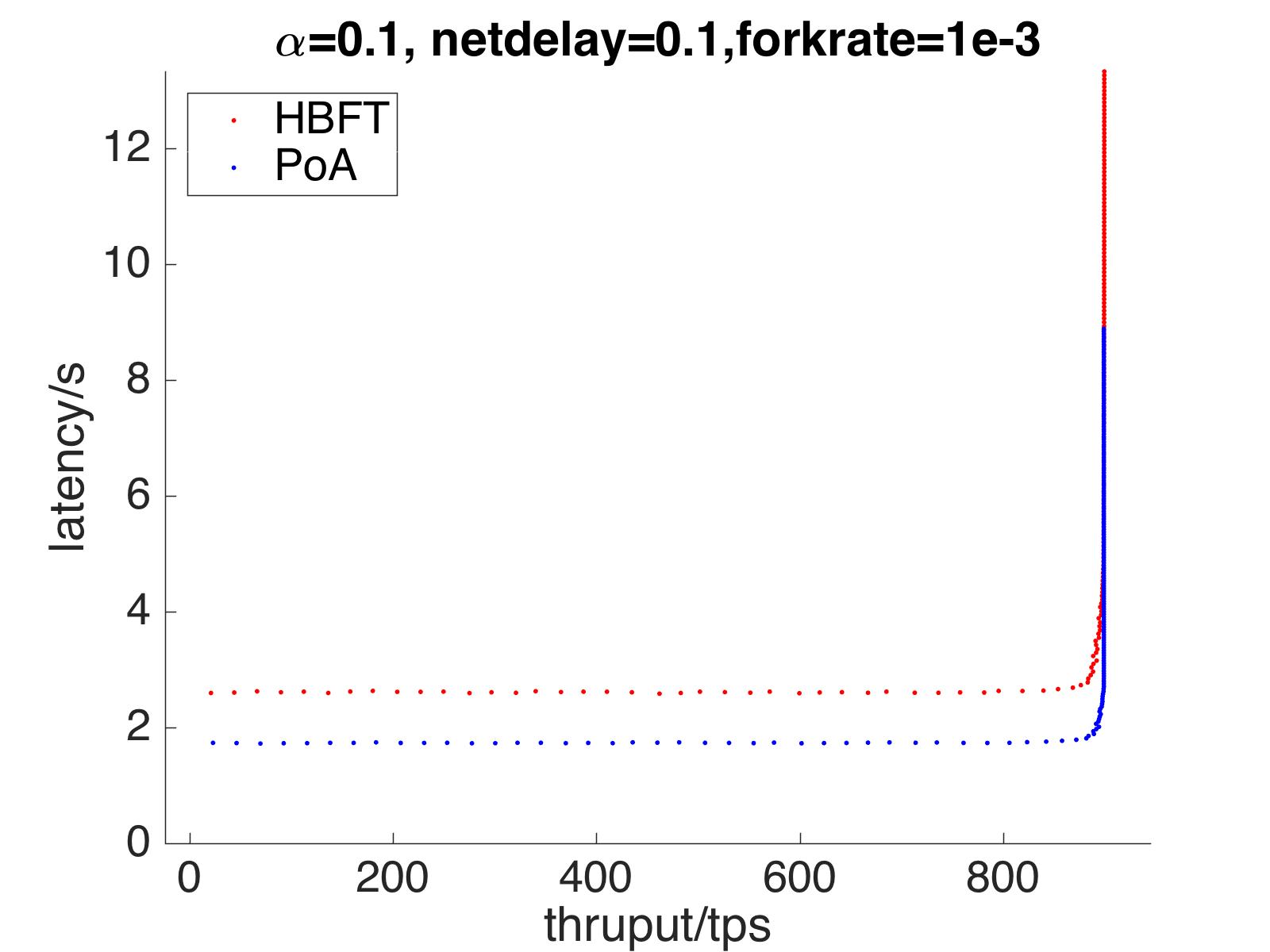}
\caption{HBFT vs. PoA throughput-latency plot across different byzantine ratio $\alpha$ and networkdelay factor $D$, and the block size range from 40 to 4000.}
\label{fig4}
\end{figure}

In the previous subsections, we showed  that the performance of PoA is dominated by HBFT in all numerical experiments. Although the
communication overhead hurts the performance of HBFT, the bandwidth wasted by forks in PoA bites on the performance of PoA. In our experiments, the minimal fork rate is $0.1$, which is high enough to  make  HBFT outperform PoA.  
In practice, if the malicious nodes use different strategies, the fork rate can vary in a wider range.  For instance, on the one hand, by introducing economic penalty,  the fork rate in  Bitcoin is only $0.0178$ according to  \cite{decker2013information} in 2013, and this fork rate can be reduced further by stronger economic penalty. 
On the other hand,  in Proof-of-Stake (PoS) algorithms like Peercoin, malicious nodes could perform Nothing-at-stakes attack to create a large number of forks \cite{coa}, resulting a large fork rate. In this work, we fix other parameters and vary the fork rate in $(0, 1)$, and find that the PoA will dominate HBFT when the fork rate is less than $0.001$. 
We show our comparison when taking the fork rate at $0.001$ in  Figure~\ref{fig4}.
By this comparison, we conclude that PoA can be a better choice if the fork rate can be pressed low by introducing some discouraging on forks.


\subsection{When the system is large } 
\begin{figure}
\centering
\includegraphics[width=8cm]{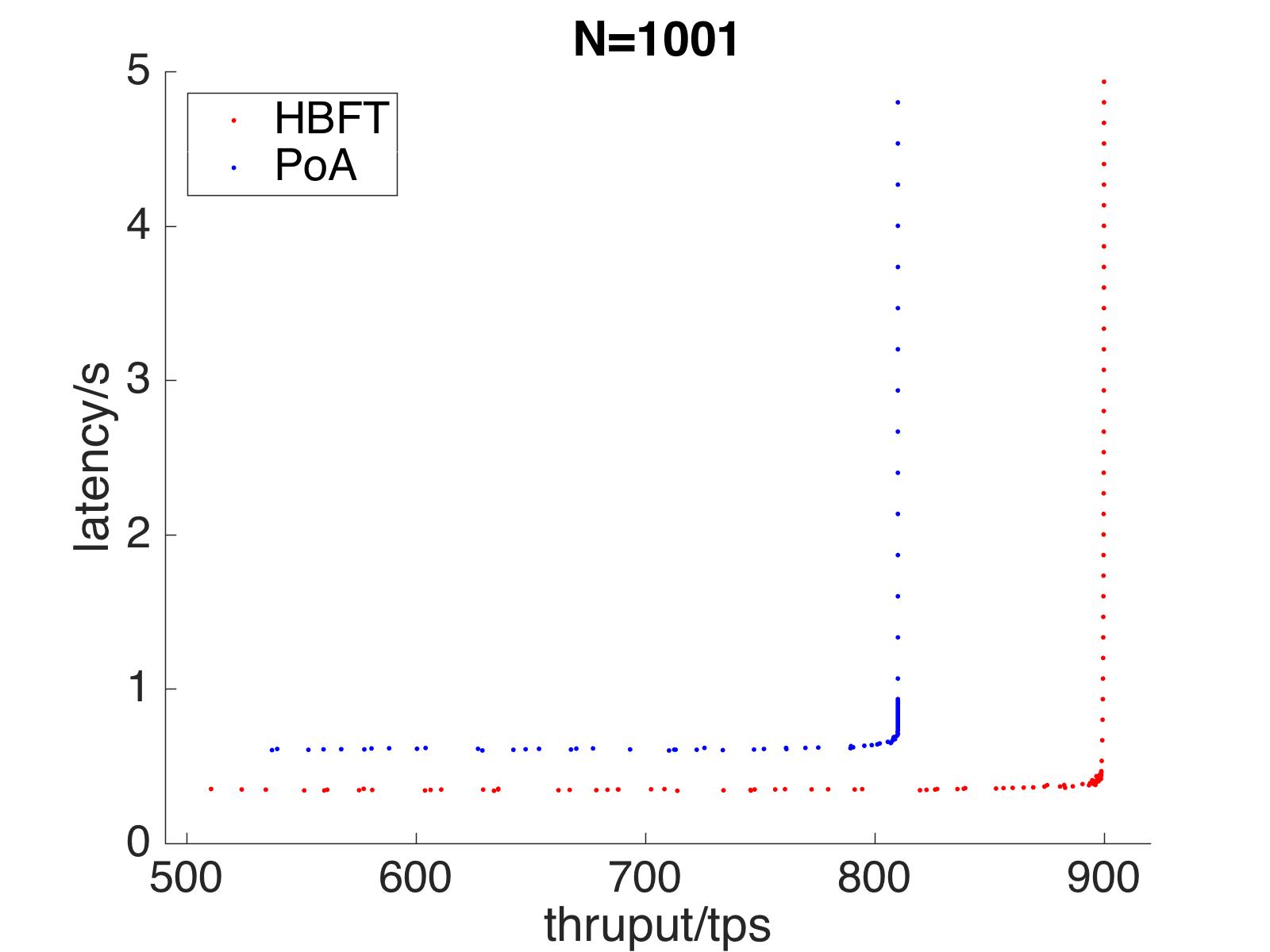}
\caption{HBFT vs. PoA throughput-latency plot with $N=1001$, byzantine ratio $\alpha=0.1$ and network delay $D=0.1$, and the block size range from 20 to 4000. HBFT no longer dominates PoA under this circumstance.}
\label{fig5}
\end{figure}

In our previous numerical experiments, the number of nodes is set to be a relatively small number $N=101$ for the convenience of our experiments. 
In practice, HBFT and PoA are deployed on much larger systems. To make sure that our comparison still makes sense in a large system, 
we set $N=1001$, and present the same comparison 
in Figure 5, which shows that the dominance of BFT over PoA still holds. Therefore, although contrary to the common belief, we are confident that 
HBFT dominates   PoA 
even in a practically larger system.

\subsection{Frontier identification}

\begin{figure}
\centering
\includegraphics[width=8cm]{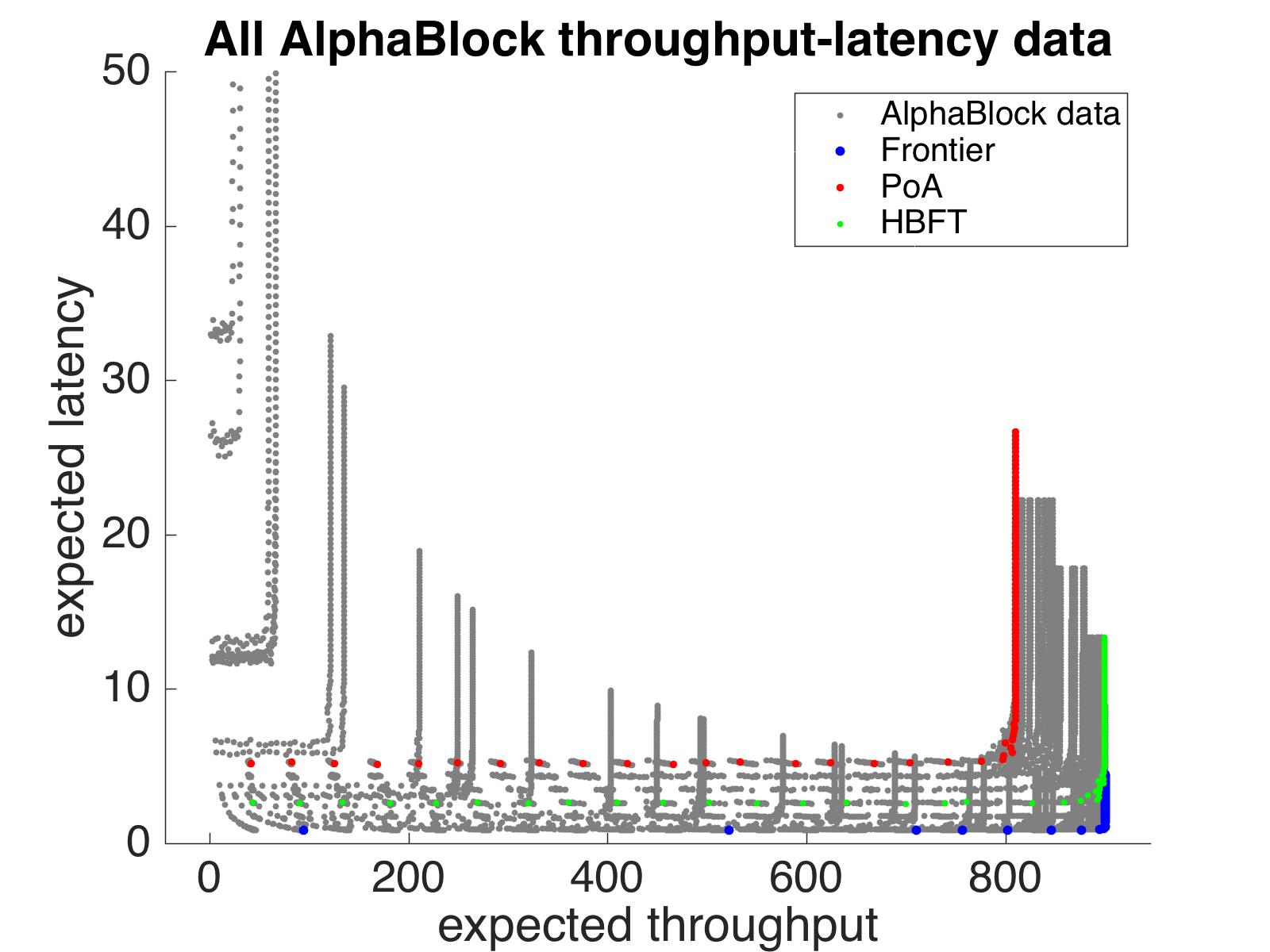}
\caption{All AlphaBlock throughput-latency plot, with HBFT, PoA and frontier points highlighted}
\label{fig5}
\end{figure}

Finally, let us zoom out of the comparison between 
HBFT and PoA, and go to find better systems in AlphaBlock
in terms of throughput and latency. The variables we can 
choose here are committee size $C$, endorsement size $d$ and block size $B_{size}$. We show in figure 6 the throughput-latency performance of AlphaBlock systems with different choices of the three variables above. In this figure, we highlight those for HBFT, PoA, and those on the Pareto frontier who has the highest throughput with a certain level of latency or the lowest latency with a certain level of throughput. It may not be possible to achieve some throughput-latency on the Pareto frontier, but we can sense
the performance of a system by its difference to the Pareto frontier. 

\section{Conclusion}

This paper proposes a general framework termed as AlphaBlock to evaluate consensus protocols for blockchain.  We compare the performance of Byzantine Fault Tolerance (BFT) and Nakamoto Consensus (NC), and we take Hotstuff-BFT (HBFT) and Proof-of-Authority (PoA) as two specific examples.
In comparison, we find that HBFT outperforms PoA in most practical settings, which contrasts the common belief.
This out-performance can be reversed if the fork rate in PoA can be reduced sufficiently by some discouragement on forks.
We also identify the Pareto-optimal choices of committee size $C$, endorsement size $d$ and block size $B_{\textrm{size}}$ in the framework of AlphaBlock, which provides a direction for improving the performance of blockchain consensus algorithms. 

\section{Appendices}
{\it Proof of Theorem \ref{thm}: }

For both protocols, when $B_{\textrm{size}} \rightarrow \infty$, we have
$B_{\textrm{interval}}\rightarrow  \frac{B_{\textrm{size}}}{B_{\textrm{width}}}$ and $C_{\textrm{rate}}\rightarrow 0$. 

Recall the definition of throughput 
\begin{equation}
T=\frac{B_{\textrm{size}}}{B_{\textrm{interval}}}\times B_{\textrm{rate}}\times B_{\textrm{eff}}. 
\end{equation}
So when $B_{\textrm{size}}\rightarrow  \infty$, we have
\begin{equation}
T\rightarrow B_{\textrm{width}}\times B_{\textrm{rate}} \times (1-F_{\textrm{rate}}).    
\end{equation}
Since the block rate is $1-\alpha$ for both protocals, we  have 
\begin{equation}
\frac{T(PoA)}{T(HBFT)}\rightarrow \frac{(1-F_{\textrm{rate}})_{PoA}}{(1-F_{\textrm{rate}})_{HBFT}}=1-\alpha,
\end{equation}
where $T(PoA)$ and $T(HBFT)$ are   the throughputs of PoA and HBFT respectively. 


\bibliographystyle{ACM-Reference-Format}
\bibliography{sample-base}


\end{document}